\documentclass[11pt]{article}
\usepackage{times}
\usepackage{geometry}
\usepackage[utf8]{inputenc}
\usepackage{amsmath}
\usepackage{multirow}
\usepackage{enumitem,amssymb}
\usepackage{graphicx}
\usepackage{authblk}
\usepackage{url}
\usepackage{soul}
\usepackage[normalem]{ulem}
\usepackage[
 colorlinks=true,    
 linkcolor=blue,     
 citecolor=blue,     
 filecolor=blue,  
 urlcolor=blue,      
 final=true
]{hyperref}
\usepackage[numbers,sort&compress]{natbib}
\usepackage{enumitem}
\usepackage[symbol]{footmisc}
\usepackage[dvipsnames]{xcolor}

\setenumerate{itemsep=0mm}

\begin{document}

\title{\large Snowmass 2021\\
\LARGE  BSM Higgs Production at a Muon Collider}
\author[a]{Tao Han}
\author[b]{Shuailong Li}
\author[b]{Shufang Su}
\author[c]{Wei Su}
\author[d,e]{Yongcheng Wu}

\affil[a]{\small Department of Physics and Astronomy, University of Pittsburgh, Pittsburgh, PA 15260, USA}
\affil[b]{\small Department of Physics, University of Arizona, Tucson, Arizona 85721, USA}
\affil[c]{\small Korea Institute for Advanced Study, Seoul 02455, Korea}
\affil[d]{\small Department of Physics, Oklahoma State University, Stillwater, OK, 74078, USA}
\affil[e]{\small Department of Physics and Institute of Theoretical Physics, Nanjing Normal University, Nanjing, 210023, China}
\maketitle

\footnotetext[1]{email address:
\href{mailto:than@pitt.edu}{than@pitt.edu},
\href{mailto:shuailongli@email.arizona.edu}{shuailongli@email.arizona.edu},
\href{mailto:shufang@arizona.edu}{shufang@arizona.edu},
\href{mailto:weisu@kias.re.kr}{weisu@kias.re.kr},
\href{mailto:ycwu0830@gmail.com}{ycwu0830@gmail.com}
}
\noindent {\large \bf Thematic Areas:}  

\noindent $\blacksquare$ (EF08) BSM: Model specific explorations \\
\noindent $\blacksquare$ (TF07) Collider phenomenology \\
\begin{abstract}
\large
The potential of the non-Standard Model heavy Higgs bosons in 2HDM at a muon collider is studied. The pair production of the non-SM Higgs bosons via the universal gauge interactions is the dominant mechanism once above the kinematic threshold. On the other hand, single Higgs boson production associated with a pair of heavy fermions is also important in the parameter region with enhanced Yukawa couplings. Both $\mu^+\mu^-$ annihilation channels and Vector Boson Fusion  processes are considered, as well as radiative return $s$-channel production.   Different types of 2HDMs can also be distinguishable for moderate and large values of $\tan\beta$.
\end{abstract}


\large

\section{ Introduction}
There have been renewed interests for muon colliders operating at high energies in the range of multi-TeV~\cite{Delahaye:2019omf,Han:2020uid,Long:2020wfp}. This would offer great physics opportunity to open  unprecedented new energy threshold for new physics, and provide precision measurements in a clean environment in leptonic collisions~\cite{Capdevilla:2021rwo,Liu:2021jyc,Huang:2021nkl,Yin:2020afe,Buttazzo:2020eyl,Capdevilla:2020qel,Han:2020pif,Han:2020uak,Costantini:2020stv}. Recent studies indeed demonstrated the impressive physics potentials exploring the electroweak sector, including precision Higgs boson coupling measurements~\cite{Han:2020pif}, the electroweak dark matter detection~\cite{Han:2020uak}, and discovery of other beyond the Standard Model (BSM) heavy particles~\cite{Costantini:2020stv}.

In this note, we summarize the results on the discovery potential for the non-SM heavy Higgs bosons at a high-energy muon collider in the framework of 2HDMs~\cite{Branco:2011iw}. We adopt the commonly studied four categories according to the assignments of a discrete $\mathbb{Z}_2$ symmetry, which dictates the pattern of the Yukawa couplings. We take a conservative approach in the alignment limit for the mixing parameter so that there are no large corrections to the SM Higgs physics.

We consider the benchmark energies for the  muon colliders  \cite{Delahaye:2019omf} in the range of $\sqrt s=3-30$ TeV, with the integrated luminosity scaled as
\begin{align}
\label{eq:luminosity}
\mathcal{L}= \left({\frac{\sqrt s}{10\ {\rm TeV}}}\right)^2 \times 10^4\ {\rm fb}^{-1}.
\end{align}
We study both the heavy Higgs boson pair production as well as single production associated with two heavy fermions. Both $\mu^+\mu^-$ annihilation channels and Vector Boson Fusion (VBF) processes are considered, which are characteristically different. We also  analyze the radiative return $s$-channel production of a heavy Higgs boson in $\mu^+\mu^-$ annihilation, given the possible enhancement of the muon Yukawa couplings in certain models. Combining together the production channels and the decay patterns, we also show how the four different types of 2HDMs can be distinguished.

\section{Higgs pair production}

\begin{figure}[!thb]
    \centering
    \includegraphics[width=0.8\textwidth]{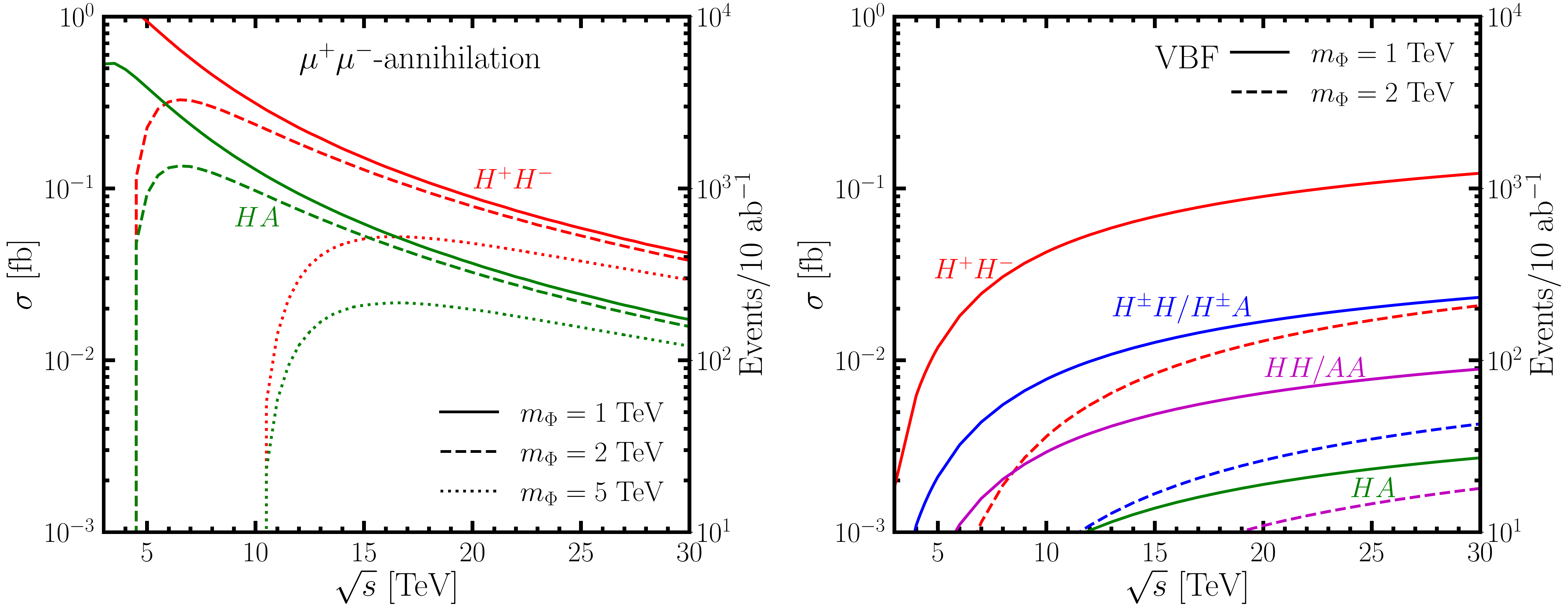}
    \caption{Pair production cross sections versus the c.m.~energy $\sqrt{s}$ for annihilation (left panel) and VBF process (right) in the alignment limit $\cos(\beta-\alpha)=0$. The vertical axis on the right shows the corresponding event yields for a 10 ${\rm ab}^{-1}$ integrated luminosity.}
    \label{fig:Pair_CS}
\end{figure}

The cross sections of the pair productions in the alignment limit $\cos(\beta-\alpha)=0$ via annihilation and VBF are shown in~\autoref{fig:Pair_CS}. For the annihilation, we see the threshold behavior for a scalar pair production in a $p$-wave as $\sigma\sim \beta^3$ with $\beta = \sqrt{1-4m_H^2/s}$. Well above the threshold, the cross section asymptotically approach $\sigma\sim \alpha^2/s$ which is insensitive to the heavy Higgs mass. The VBF processes become increasingly important at high c.m.~energies. We see the logarithmic enhancement over the energy $\log^2(s/m_\mu^2)$ (or $\log^2(s/m_V^2)$). Unlike the production via annihilation, the cross section for the VBF processes are very sensitive to the heavy Higgs masses.

In general, the annihilation process yields more events than the VBF process, except for the $H^+H^-$ production. Different decay channels of the scalars, which have different branching fractions in different types of 2HDM, can be used to help distinguishing four different types. The leading signal channels for different cases are listed in~\autoref{tab:pair} and several observations can be made:
\begin{itemize}
    \item{} For low values of $\tan\beta <5$, the four models cannot be distinguished since the dominating decay channels are the same: $H/A \rightarrow t\bar{t}$, $H^\pm \rightarrow tb$.
    \item{} For large values of $\tan\beta>10$,
    the decay modes of $H/A \rightarrow \tau^+\tau^-$, $H^\pm \rightarrow \tau\nu$ become substantial in Type-L, which can be used to separate Type-L from the other three types of 2HDMs.
    \item{}  For $\tan\beta>5$, the enhancement of the bottom Yukawa coupling with $\tan\beta$ in Type-II/F leads to the growing and even the leading of $H/A\to b\bar b$ decay branching fraction, which can be used to separate Type-II/F from the Type-I 2HDM.
    \item{}Type-II and Type-F cannot be distinguished for all ranges of $\tan\beta$ based on the leading channel, since the leptonic decay mode is always sub-dominate comparing to decays into top or bottom quarks in all ranges of $\tan\beta$.    The full discrimination is only possible at $\tan\beta>10$ if the sub-leading $H^\pm\to \tau\nu$ and $H/A\to \tau^+\tau^-$ decays in Type-II can be detected, which has a
    branching fraction about $10\%$.
\end{itemize}

\begin{table}[!bht]
    \centering
    \begin{tabular}{|c|c|c|c|c|c|}\hline
         &  production &Type-I & Type-II & Type-F & Type-L \\ \hline
         \multirow{3}{*}{small $\tan\beta<5$}&$H^+H^-$&\multicolumn{4}{c|}
         {$t\bar b, \bar t b$} \\
          &$HA/HH/AA$&\multicolumn{4}{c|}{$t\bar t, t\bar t$} \\
          &$H^\pm H/A$&\multicolumn{4}{c|}{$tb, t\bar t$} \\  \hline
          \multirow{4}{*}{intermediate $\tan\beta$}&$H^+H^-$&\multicolumn{3}{c|}{$t\bar b, \bar t b$}&$tb, \tau\nu_\tau$ \\  \cline{3-5}
         &$HA/HH/AA$&$t\bar t, t\bar t$&\multicolumn{2}{c|}{$t\bar t, b\bar b$}&$t\bar t,\tau^+\tau^-$\\
          &$H^\pm H/A$&$tb, t\bar t$&\multicolumn{2}{c|}{$tb, t\bar t$;\  $tb,b\bar b$}&$tb, t\bar t$;\ $tb,\tau^+\tau^-$; \\
          & & &  \multicolumn{2}{c|}{}& $\tau\nu_\tau, t\bar t$;\ \  $\tau\nu_\tau, \tau^+\tau^-$   \\ \hline
         \multirow{3}{*}{large $\tan\beta>10$}&$H^+H^-$& {$t\bar b, \bar t b$}& $tb, t b(\tau\nu_\tau)$&{$t\bar b, \bar t b$}&$\tau^+\nu_\tau, \tau^-  \nu_\tau$\\
         &$HA/HH/AA$&$t\bar t, t\bar t$&$b\bar{b},b\bar{b}(\tau^+\tau^-)$& {$b\bar b, b\bar b$}&$\tau^+\tau^-, \tau^+ \tau^-$\\
         &$H^\pm H/A$&$tb, t\bar t$&$tb(\tau\nu_\tau),b\bar b(\tau^+\tau^-)$& $tb,b\bar b$&$\tau^\pm  \nu_\tau,\tau^+ \tau^- $\\ \hline
    \end{tabular}
    \caption{Leading signal channels of Higgs pair production for various 2HDMs in different regions of small, intermediate and large $\tan\beta$. Channels in the parenthesis are the sub-leading channels. }
    \label{tab:pair}
\end{table}

\begin{figure}[!tb]
    \centering
    \includegraphics[width=0.8\textwidth]{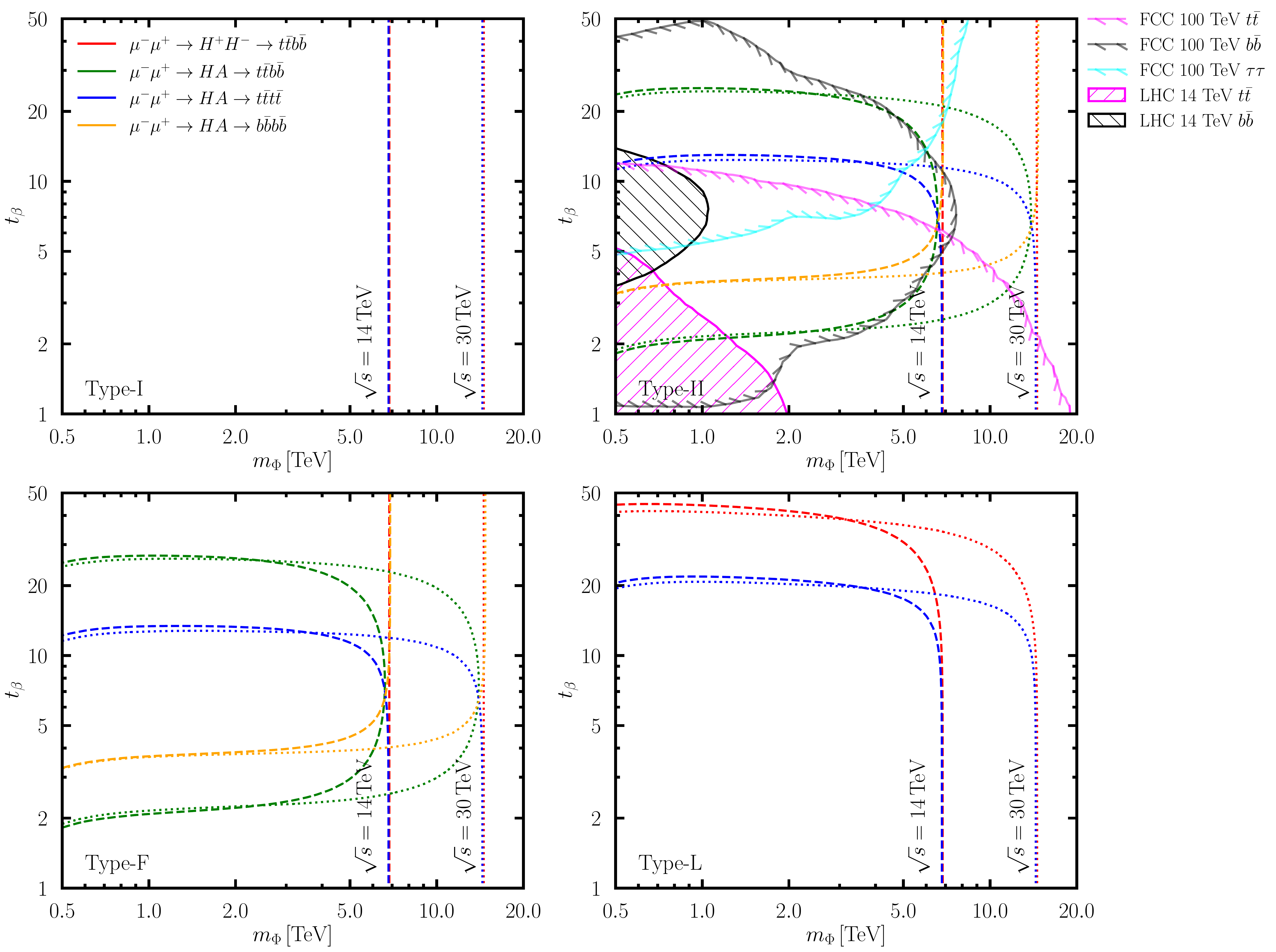}
    \caption{95\% C.L. exclusion contour at muon collider with center of mass energy $\sqrt{s}=14$ (dash curves), 30 (dotted curves) TeV for different types of 2HDM from pair production channels with annihilation contribution only. For the Type-II 2HDM, the 95\% C.L. exclusion limits from the High-Luminosity Large Hadron Collider (HL-LHC) with 3 ab$^{-1}$ as well as the 100 TeV $pp$ collider with 30 ab$^{-1}$ are also shown (taken from  Ref.~\cite{Craig:2016ygr}).}
    \label{fig:95reach}
\end{figure}

The reach of the heavy Higgs pair production via annihilation at a muon collider is also summarized in~\autoref{fig:95reach} with the comparison of hadron collider reach in the Type-II.

\section{Higgs boson associated production with a pair of heavy fermions}

\begin{figure}[!thb]
    \centering
    \includegraphics[width=0.8\textwidth]{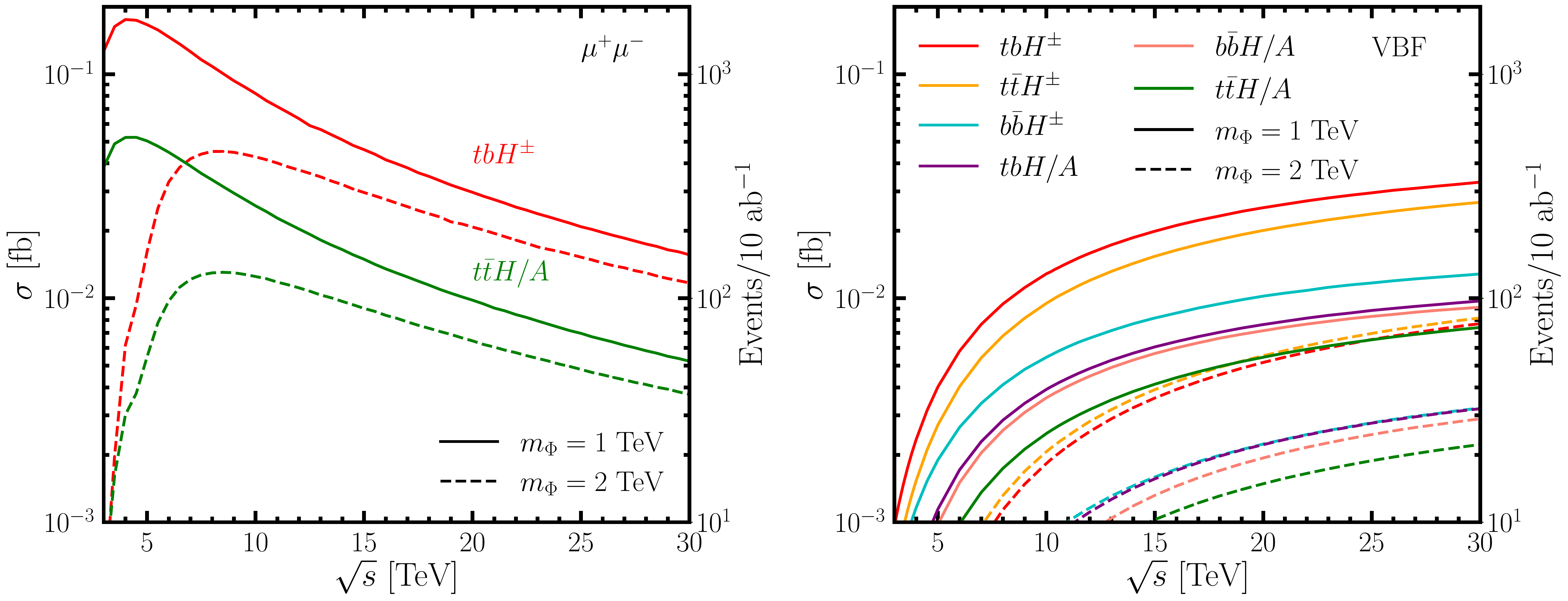}
    \caption{Cross sections versus the c.m.~energy $\sqrt{s}$ for a single heavy Higgs production associated with a pair of fermions via annihilation (left) and VBF (right) for $t_\beta=1$. Acceptance cuts  of $p_{T}^f>50$ GeV and  $10^\circ<\theta_f<170^\circ$ are imposed on all outgoing fermions. A veto cut of $0.8m_\Phi<m_{f f^\prime}<1.2m_\Phi$ is applied to the associated fermions to remove contributions from resonant Higgs decays.   The vertical axis on the right shows the corresponding event yields for a 10 ab$^{-1}$ integrated luminosity.}
    \label{fig:com_ffH}
\end{figure}

Heavy Higgs bosons can also be abundantly produced in association with a pair of heavy fermions at a muon collider. The cross sections of such processes via annihilation or VBF for $t_\beta = 1$ and $\cos({\beta-\alpha})=0$ are shown in~\autoref{fig:com_ffH}. Unlike the pair production case,  the production cross sections for a single heavy Higgs boson produced in association with a pair of heavy fermions depend sensitively on $\tan\beta$.   Combining with different decay channels of the heavy scalar, we can also try to distinguish different types of 2HDM.

In~\autoref{tab:fermion} we summarized the leading signal channels of the Higgs associated production with fermions in four types of 2HDMs in different regimes of $\tan\beta$.  Several observations can be made:
\begin{itemize}
    \item In the small $\tan\beta<5$ region, all six production channels have sizable production cross sections.  However, it is hard to distinguish different types of 2HDMs since they all lead to the same final states.
    \item In the large $\tan\beta>10$ region, all the production channels for the Type-I are suppressed, while Type-II/F have sizable production in $tbH^\pm$, $bbH^\pm$, $bbH/A$ and $tbH/A$ channels.  Type-II and Type-F can be further separated by studying the sub-dominant decay channels of $H^\pm \rightarrow \tau \nu_\tau$ and $H/A \rightarrow \tau^+\tau^-$ in the Type-II.   Same final states of Type-II can also be obtained via $\tau\nu_\tau H^\pm$, $\tau^+\tau^- H^\pm$, $\tau^+\tau^- H/A$ and $\tau\nu_\tau H/A$ production.
    \item The intermediate range of $\tan\beta$ is the most difficult
    region for all types of 2HDMs, since top Yukawa couplings are reduced, while bottom Yukawa coupling is not big enough to compensate, resulting in a rather low signal production rate.  A rich set of final states, however, are available given the various competing decay modes of $H^\pm$ and $H/A$.
    \item At very large value of $\tan\beta>50$, the tau-associated production $\tau\nu_\tau H^\pm$, $\tau^+\tau^- H^\pm$, $\tau^+\tau^- H/A$ and $\tau\nu_\tau H/A$ would be sizable for Type-L.
\end{itemize}

\begin{table}[tb]
    \centering
    \begin{tabular}{|c|c|c|c|c|c|}
        \hline
         &  production &Type-I & Type-II & Type-F & Type-L \\
         \hline
         \multirow{6}{*}{small $\tan\beta<5$} & $tbH^\pm$    & \multicolumn{4}{c|}{$tb, tb$} \\
         & $t\bar{t}H^\pm$   & \multicolumn{4}{c|}{$t\bar{t}, tb$} \\
         & $b\bar{b}H^\pm$   & \multicolumn{4}{c|}{$b\bar{b}, tb$} \\
        &$t\bar{t}H/A$ & \multicolumn{4}{c|}{$t\bar t, t \bar{t}$}\\
        &$b\bar{b}H/A$ & \multicolumn{4}{c|}{$b\bar b, t \bar{t}$}\\
        &$tbH/A$     & \multicolumn{4}{c|}{$tb, t \bar{t}$}\\               \hline
          \multirow{6}{*}{intermediate $\tan\beta$} &$tbH^\pm$ &\multicolumn{3}{c|}{$tb,tb$}&$tb,tb$; $tb,\tau\nu_\tau$ \\
          \cline{3-6}
           &$t\bar{t}H^\pm$ &\multicolumn{3}{c|}{$t\bar{t},tb$}&$t\bar{t},tb$; $t\bar{t},\tau\nu_\tau$ \\
          \cline{3-6}
            &$b\bar{b}H^\pm$ &\multicolumn{3}{c|}{$b\bar{b},tb$}&$b\bar{b},tb$; $b\bar{b},\tau\nu_\tau$ \\
          \cline{3-6}
                                                    &$t\bar{t}H/A$& $t\bar{t}, t\bar{t}$&\multicolumn{2}{c|}{$t\bar{t},t\bar{t}$; $t\bar{t}, b\bar{b}$} &$t\bar{t}, t\bar{t}$; $t\bar{t},\tau^+\tau^-$ \\
         &$b\bar{b}H/A$& $b\bar{b}, t\bar{t}$&\multicolumn{2}{c|}{$b\bar{b}, t\bar{t}$; $b\bar{b}, b\bar{b}$} &$b\bar{b}, t\bar{t}$; $b\bar{b},\tau^+\tau^-$\\
         &$tbH/A$& $tb,t\bar{t}$&\multicolumn{2}{c|}{$tb,t\bar{t}$; $tb,b\bar{b}$} &$tb,t\bar{t}$; $tb,\tau^+\tau^-$ \\
        \hline
          \multirow{4}{*}{large  $\tan\beta>10$}& $tbH^\pm$ & $-$ & {$tb,tb (\tau\nu_\tau)$} & {$tb,tb$} & $-$\\
          \cline{4-5}
          & $bbH^\pm$ &$-$ & {$bb,tb (\tau\nu_\tau)$} & {$bb,tb$} & $-$\\
          \cline{4-5}
          &$b\bar{b}H/A$&$-$ & {$b\bar{b}, b\bar{b} (\tau^+\tau^-)$}&{$b\bar{b}, b\bar{b}$}& $-$\\
          \cline{4-5}
          &$t\bar{b}H/A$&$-$ & {$t\bar{b}, b\bar{b} (\tau^+\tau^-)$}&{$t\bar{b}, b\bar{b}$}& $-$\\
          \hline
\multirow{4}{*}{very large  $\tan\beta>50$}
    &$\tau\nu_\tau H^\pm$& \multicolumn{3}{c|}{$-$}& $\tau\nu_\tau, \tau\nu_\tau$\\
    &$\tau^+\tau^- H^\pm$& \multicolumn{3}{c|}{$-$}& $\tau^+\tau^-, \tau\nu_\tau$\\
    &$\tau^+\tau^- H/A$& \multicolumn{3}{c|}{$-$}& $\tau^+\tau^-, \tau^+\tau^-$\\
     &$\tau\nu_\tau H/A$& \multicolumn{3}{c|}{$-$}& $\tau \nu_\tau, \tau^+\tau^-$\\
        \hline
    \end{tabular}
    \caption{Leading signal channels of single Higgs associated production with a pair of fermions for various 2HDMs in different regions of small, intermediate and large $\tan\beta$.  Channels in the parenthesis are the sub-leading channels.
    }
    \label{tab:fermion}
\end{table}

\section{Radiative return}

While the cross sections for heavy Higgs pair production are un-suppressed under the alignment limit, the cross section has a threshold cutoff at $m_\Phi\sim \sqrt{s}/2$.   The resonant production for a single heavy Higgs boson may  further extend the coverage to about
$m_\Phi\sim \sqrt{s}$, as long as the coupling strength to $\mu^+\mu^-$ is big enough. The drawback for the resonant production is that the collider energy would have to be tuned close to the mass of the heavy Higgs, which is less feasible at future muon colliders. A promising mechanism is to take advantage of the initial state radiation (ISR), so that the colliding energy is reduced to a lower value for a  resonant production, thus dubbed the ``radiative return''~\cite{Chakrabarty:2014pja}.

\begin{figure}[!tb]
    \centering
    \includegraphics[width=0.8\textwidth]{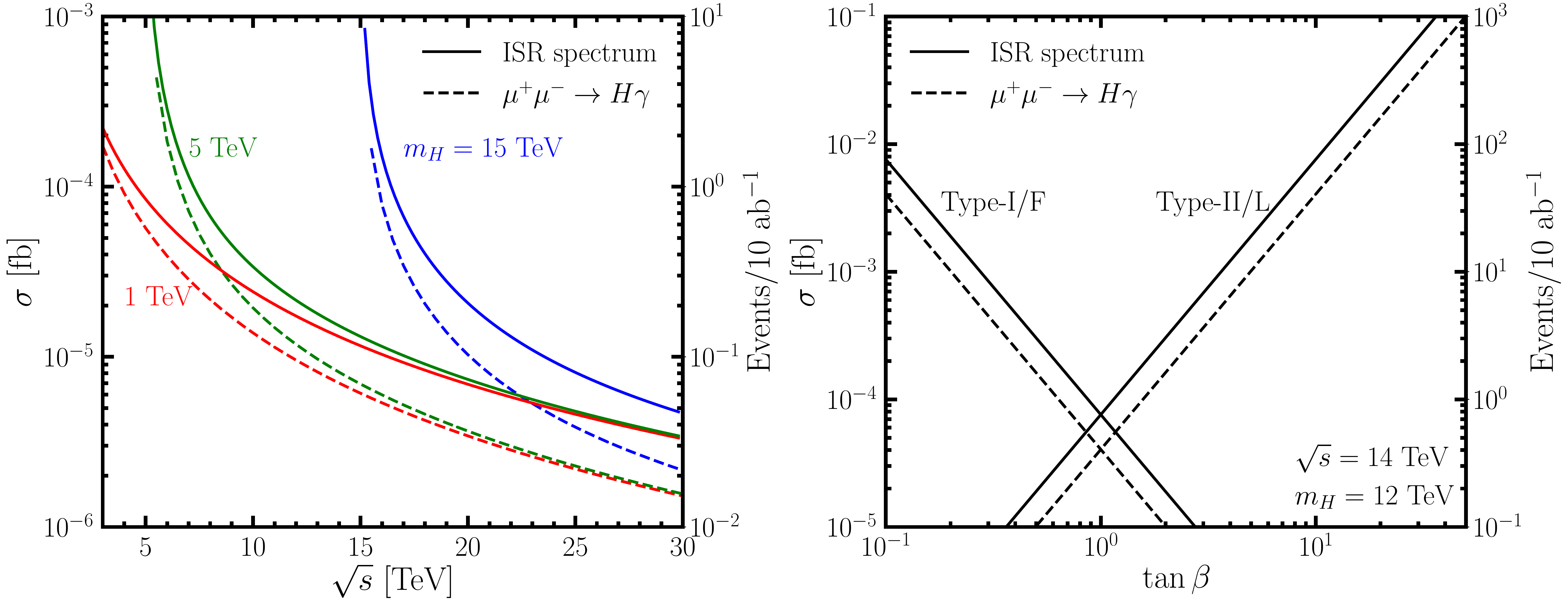}
    \caption{Cross sections of single heavy Higgs $H$ production through the radiative return. Left panel is for $m_H=1$, 2 and 15 TeV and $\tan\beta=1$ versus the c.m.~energy $\sqrt s$, with the solid curves for the convoluted ISR spectrum and the dashed curves for single photon radiation $\mu^+\mu^- \rightarrow H \gamma$ with $10^\circ<\theta_\gamma<170^\circ$.  Right panel is for the $\tan\beta$ dependence of the cross section for $\sqrt{s}=14$ TeV and $m_H=12$ TeV. The vertical axis on the right shows the corresponding event yields for a 10 ab$^{-1}$ integrated luminosity.}
    \label{fig:gammaH}
\end{figure}

The cross section of the ``radiative return'' process is calculated in two ways: (1) $\mu^+\mu^-\to \gamma H$ and (2) $\mu^+\mu^-\to H$ with ISR spectrum. The results are given in~\autoref{fig:gammaH}, where the left panel shows the dependence on c.m.~energy while the right panel shows the $\tan\beta$ dependence. We can see that the cross section of such process increases as the heavy Higgs mass approaches the collider c.m.~energy. On the other hand, although, the cross section is much smaller than the other productions channels in previous sections at moderate $\tan\beta$, it could be largely enhanced at large (small) $\tan\beta$ for Type-II/L (Type-I/F). It could even be the dominant production for heavy Higgs in the large $\tan\beta$ region of Type-L when pair production is kinematically forbidden and quark associated productions are suppressed.

\section{Executive summary}
Motivated by the recent interests in future high energy muon colliders, we explore the physics reach in studying heavy Higgs bosons in 2HDM at such facilities.
We point out that the pair production of the non-SM Higgs bosons via the universal gauge interactions is the dominant mechanism once above the kinematic threshold. In contrast, single Higgs boson production associated with a pair of heavy fermions is important in the parameter region with enhanced Yukawa couplings. As such, $\mu^+\mu^-$ annihilation channels and Vector Boson Fusion  processes are complementary for discovery and detailed property studies.
The radiative return mechanism in the $s$-channel production can extend the mass coverage close to the collider c.m.~energy. Different types of 2HDMs can also be distinguishable for moderate and large values of $\tan\beta$.

\bibliographystyle{JHEP}
\bibliography{references}

\providecommand{\href}[2]{#2}\begingroup\raggedright\begin{thebibliography}{10}

\bibitem{Delahaye:2019omf}
J.~P. Delahaye, M.~Diemoz, K.~Long, B.~Mansouli\'e, N.~Pastrone, L.~Rivkin,
  D.~Schulte, A.~Skrinsky, and A.~Wulzer, {\it {Muon Colliders}},
  \href{http://arxiv.org/abs/1901.06150}{{\tt arXiv:1901.06150}}.

\bibitem{Han:2020uid}
T.~Han, Y.~Ma, and K.~Xie, {\it {High Energy Leptonic Collisions and
  Electroweak Parton Distribution Functions}},
  \href{http://arxiv.org/abs/2007.14300}{{\tt arXiv:2007.14300}}.

\bibitem{Long:2020wfp}
K.~Long, D.~Lucchesi, M.~Palmer, N.~Pastrone, D.~Schulte, and V.~Shiltsev, {\it
  {Muon Colliders: Opening New Horizons for Particle Physics}},
  \href{http://arxiv.org/abs/2007.15684}{{\tt arXiv:2007.15684}}.

\bibitem{Capdevilla:2021rwo}
R.~Capdevilla, D.~Curtin, Y.~Kahn, and G.~Krnjaic, {\it {A No-Lose Theorem for
  Discovering the New Physics of $(g-2)_\mu$ at Muon Colliders}},
  \href{http://arxiv.org/abs/2101.10334}{{\tt arXiv:2101.10334}}.

\bibitem{Liu:2021jyc}
W.~Liu and K.-P. Xie, {\it {Probing electroweak phase transition with multi-TeV
  muon colliders and gravitational waves}},
  \href{http://arxiv.org/abs/2101.10469}{{\tt arXiv:2101.10469}}.

\bibitem{Huang:2021nkl}
G.-y. Huang, F.~S. Queiroz, and W.~Rodejohann, {\it {Gauged
  $L^{}_{\mu}{-}L^{}_{\tau}$ at a muon collider}},
  \href{http://arxiv.org/abs/2101.04956}{{\tt arXiv:2101.04956}}.

\bibitem{Yin:2020afe}
W.~Yin and M.~Yamaguchi, {\it {Muon $g-2$ at multi-TeV muon collider}},
  \href{http://arxiv.org/abs/2012.03928}{{\tt arXiv:2012.03928}}.

\bibitem{Buttazzo:2020eyl}
D.~Buttazzo and P.~Paradisi, {\it {Probing the muon g-2 anomaly at a Muon
  Collider}},  \href{http://arxiv.org/abs/2012.02769}{{\tt arXiv:2012.02769}}.

\bibitem{Capdevilla:2020qel}
R.~Capdevilla, D.~Curtin, Y.~Kahn, and G.~Krnjaic, {\it {A Guaranteed Discovery
  at Future Muon Colliders}},  \href{http://arxiv.org/abs/2006.16277}{{\tt
  arXiv:2006.16277}}.

\bibitem{Han:2020pif}
T.~Han, D.~Liu, I.~Low, and X.~Wang, {\it {Electroweak Couplings of the Higgs
  Boson at a Multi-TeV Muon Collider}},
  \href{http://arxiv.org/abs/2008.12204}{{\tt arXiv:2008.12204}}.

\bibitem{Han:2020uak}
T.~Han, Z.~Liu, L.-T. Wang, and X.~Wang, {\it {WIMPs at High Energy Muon
  Colliders}},  \href{http://arxiv.org/abs/2009.11287}{{\tt arXiv:2009.11287}}.

\bibitem{Costantini:2020stv}
A.~Costantini, F.~De~Lillo, F.~Maltoni, L.~Mantani, O.~Mattelaer, R.~Ruiz, and
  X.~Zhao, {\it {Vector boson fusion at multi-TeV muon colliders}},  5, 2020.
\newblock \href{http://arxiv.org/abs/2005.10289}{{\tt arXiv:2005.10289}}.

\bibitem{Branco:2011iw}
G.~C. Branco, P.~M. Ferreira, L.~Lavoura, M.~N. Rebelo, M.~Sher, and J.~P.
  Silva, {\it {Theory and phenomenology of two-Higgs-doublet models}},  {\em
  Phys. Rept.} {\bf 516} (2012) 1--102,
  [\href{http://arxiv.org/abs/1106.0034}{{\tt arXiv:1106.0034}}].

\bibitem{Craig:2016ygr}
N.~Craig, J.~Hajer, Y.-Y. Li, T.~Liu, and H.~Zhang, {\it {Heavy Higgs bosons at
  low $\tan \beta$: from the LHC to 100 TeV}},  {\em JHEP} {\bf 01} (2017) 018,
  [\href{http://arxiv.org/abs/1605.08744}{{\tt arXiv:1605.08744}}].

\bibitem{Chakrabarty:2014pja}
N.~Chakrabarty, T.~Han, Z.~Liu, and B.~Mukhopadhyaya, {\it {Radiative Return
  for Heavy Higgs Boson at a Muon Collider}},  {\em Phys. Rev. D} {\bf 91}
  (2015), no.~1 015008, [\href{http://arxiv.org/abs/1408.5912}{{\tt
  arXiv:1408.5912}}].

\end{thebibliography}\endgroup

\end{document}